\documentclass[graybox]{svmult}

\usepackage{mathptmx}       
\usepackage{helvet}         
\usepackage{courier}        
\usepackage{type1cm}        
\usepackage{amsmath}
\usepackage{multirow}
\usepackage{makeidx}         
\usepackage{graphicx}        
\usepackage{multicol}        
\usepackage[bottom]{footmisc}

\usepackage{hyperref} 

\makeindex             

\newcommand{\Mjup}{$M_{\text{Jup}}$}

\newcommand {\micron}{\mbox{$\mu$m}}
\newcommand\arcsec{$^{\prime\prime}$}

\begin{document}

\title*{Adaptive optics in high-contrast imaging}
\author{Milli Julien, Mawet Dimitri, Mouillet David, Kasper Markus, Girard Julien}
\institute{Julien Milli, Julien H. Girard \at ESO,  Alonso de C\'ordova 3107, Vitacura, Santiago, Chile, \email{jmilli@eso.org}
\and Dimitri Mawet \at Caltech, 1200 E. California Blvd, Pasadena, CA 91125, USA, \email{dmawet@astro.caltech.edu}
\and Markus Kasper \at ESO, Karl-Schwarzschild-Stra{\ss}e 2, 85748 Garching, Germany, \email{mkasper@eso.org}
\and David Mouillet \at IPAG, BP 53, F-38041, Grenoble C\'edex 9, France, \email{david.mouillet@obs-ujf.grenoble.fr}}
%
%
\maketitle


\abstract{The development of adaptive optics (AO) played a major role in modern astronomy over the last three decades. By compensating for the atmospheric turbulence, these systems enable to reach the diffraction limit on large telescopes. In this review, we will focus on high contrast applications of adaptive optics, namely, imaging the close vicinity of bright stellar objects and revealing regions otherwise hidden within the turbulent halo of the atmosphere to look for objects  with a contrast ratio lower than $10^{-4}$ with respect to the central star. Such high-contrast AO-corrected observations have led to fundamental results in our current understanding of planetary formation and evolution as well as stellar evolution. AO systems equipped three generations of instruments, from the first pioneering experiments in the nineties, to the first wave of instruments on 8m-class telescopes in the years 2000, and finally to the extreme AO systems that have recently started operations. Along with high-contrast techniques, AO enables to reveal the circumstellar environment: massive protoplanetary disks featuring spiral arms, gaps or other asymetries hinting at on-going planet formation, young giant planets shining in thermal emission, or tenuous debris disks and micron-sized dust leftover from collisions in massive asteroid-belt analogs. After introducing the science case and technical requirements, we will review the architecture of standard and extreme AO systems, before presenting a few selected science highlights obtained with recent AO instruments.}

\section{Introduction}

\subsection{Science case}

\begin{figure}[b]
\includegraphics[width=\textwidth]{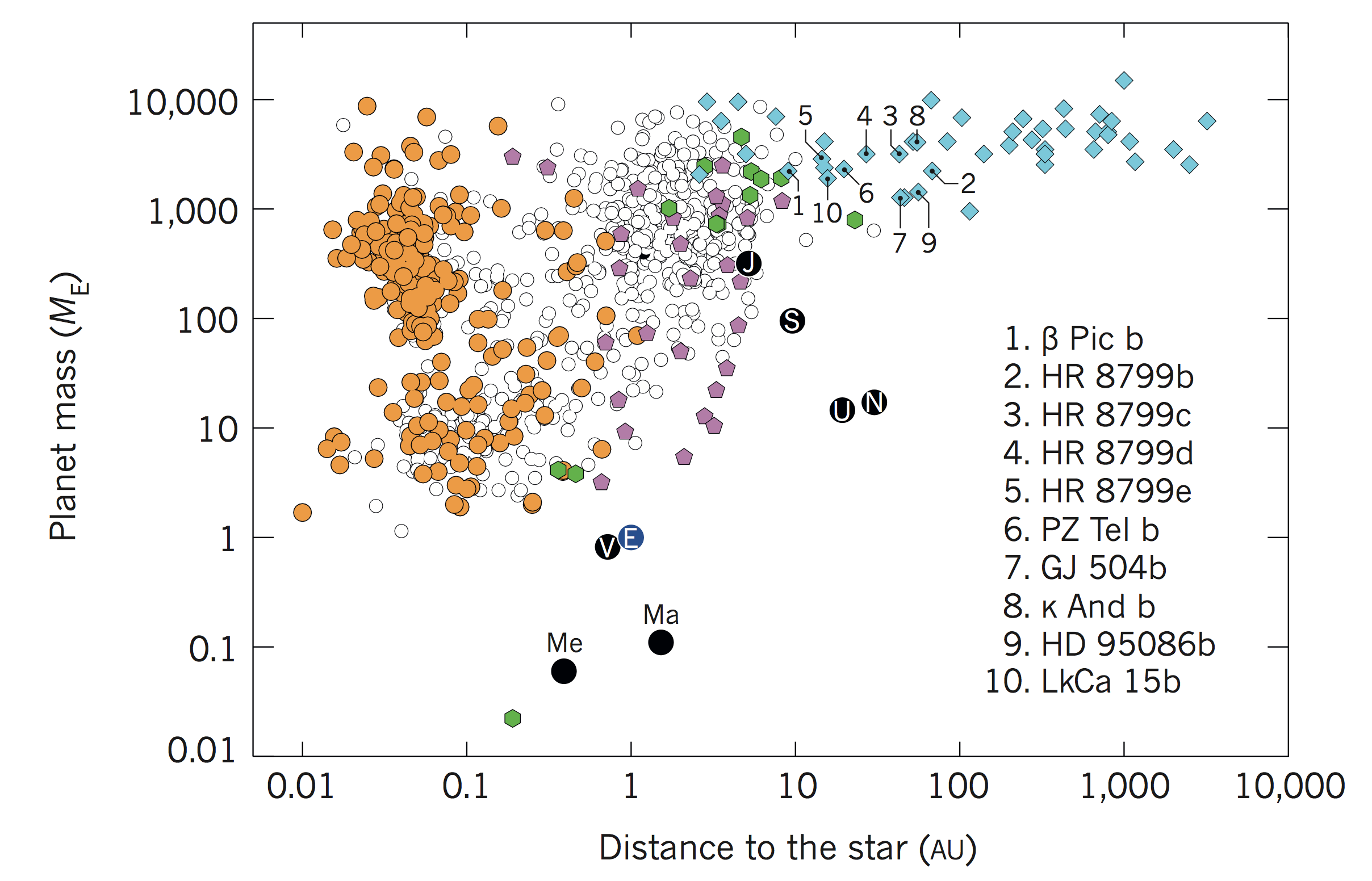}
\caption{Mass as a function of semi-major axis for solar system planets (filled black circles) and detected extrasolar planets (other symbols), from \cite{Pepe2014}. The colors indicate the detection technique: light blue for direct imaging, white for radial velocities, yellow for transits (planets with measured mass), pink for microlensing, green for pulsation timing. The 10 labelled planets are giant planets detected with AO-assisted direct imaging within 100 au of their host star and with a mass ratio to their host star below 0.02.}
\label{fig_mass_sma}       
\end{figure}

While most confirmed exoplanets were discovered by indirect techniques such as radial velocities or transits, AO-assisted direct imaging is a very rich and complementary method that can reveal the orbital motion of the planet, the spectro-photometry of its atmosphere but also the architecture and properties of its circumstellar environment. It can unveil possible interactions with a disc, whether a proto-planetary disc in case of on-going planetary accretion, or a debris disc for more evolved, gas-poor systems. 
From a statistical point of view, it probes a region, in the mass versus semi-major axis discovery space, different from other techniques, as illustrated in Fig. \ref{fig_mass_sma}. Reaching a uniform sampling of such a parameter space is essential to derive the frequency of planets as a function of mass and semi-major axis, and therefore constrain the planet formation mechanisms. For instance, theories of planet formations predict a higher efficiency of giant planet formation close to the snow line, where radial velocity and transit techniques are poorly sensitive. They require additional ingredients such as migrations and orbital instabilities to explain the current view depicted in Fig. \ref{fig_mass_sma}.

\subsection{Requirements}
\label{sec_requirements}

Imaging extrasolar planets and discs requires dedicated instruments and strategies to overcome two main challenges: 
\begin{enumerate}
\item the tiny angular separation between the star and the planet or disc. The projected separation is below 0.1\arcsec for a planet orbiting at 10 au from a star distant of 100pc.
\item the contrast between a star and its planet ranges between $10^{-4}$ for a young giant planet to $10^{-10}$ for an Earth shining in reflected light. Discs are also very tenuous, with contrasts\footnote{For extended structures, the contrast is defined per resolution elements.} ranging from $10^{-4}$ for the brightest debris discs down to  $10^{-10}$ for a an analog of our zodiacal belt at 10pc \cite{Schneider2014_AFTA}.
\end{enumerate}

\begin{figure}[b]
\includegraphics[width=\textwidth]{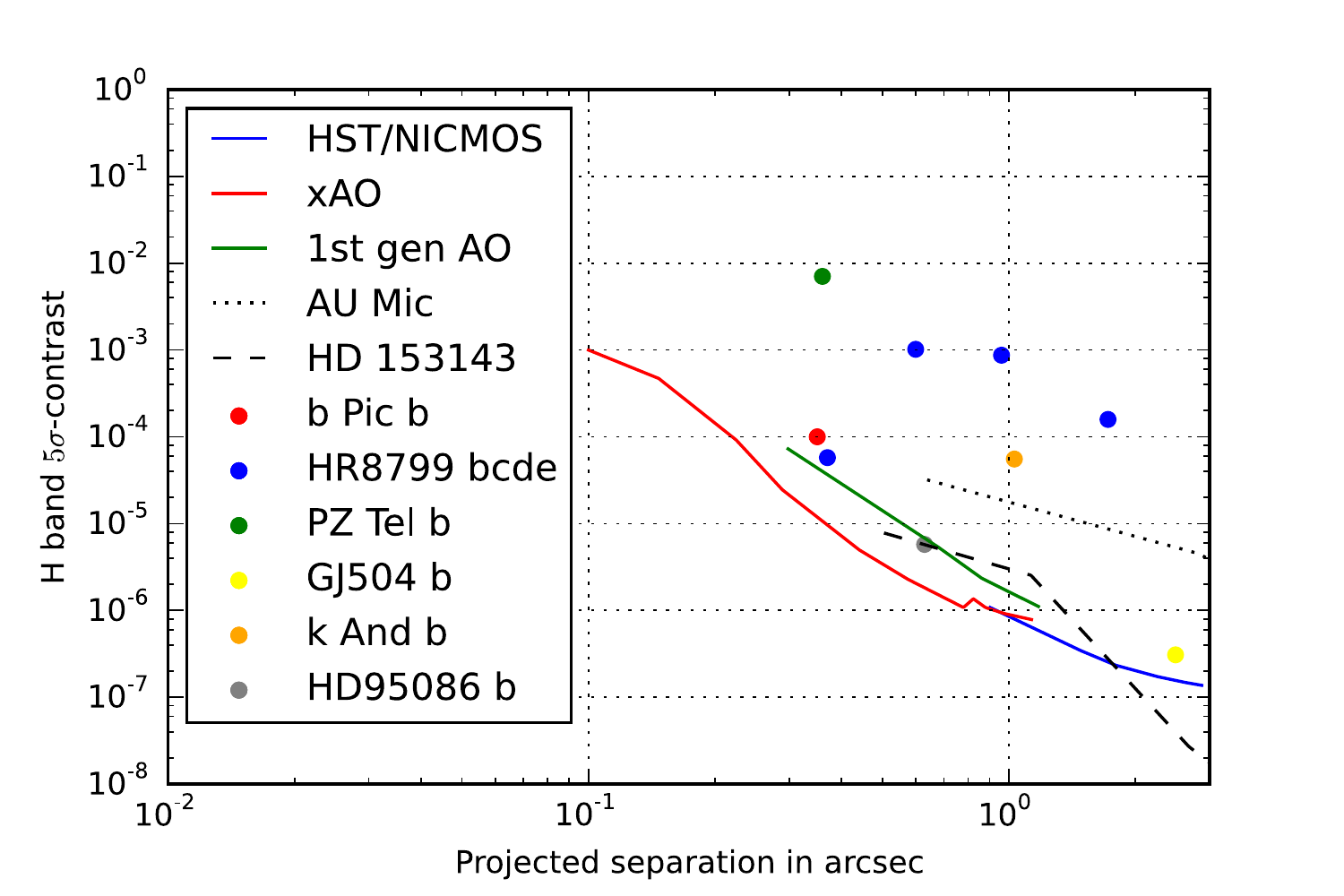}
\caption{Typical contrast obtained on sky in the H band from the first generation of AO systems (NaCo, NIRC2, NICI, HiCIAO, green curve), the second generation of extreme AO system (SPHERE/GPI, red curve), compared to the space-based instrument HST/NICMOS (blue curve). These detection curves are at $5\sigma$ for a typical 30min observations. We overplotted the typical contrast of confirmed exoplanets as well as two disk contrast expressed per resolution element (black dotted and dashed curves). }
\label{fig_contrast_vs_sep_ins}       
\end{figure}

These two requirements are summarised in Fig. \ref{fig_contrast_vs_sep_ins} that shows the separation and contrast of a few planets detected in direct imaging (dots), and two debris discs surface brightnesses (black lines). These two requirements are indissociable. From the ground, adaptive optics is one answer to the first requirement. Other techniques exist such as speckle imaging \cite{Labeyrie1970}, lucky imaging \cite{Law2006}, sparse aperture masking \cite{Baldwin1986}, interferometry \cite{Lebouquin2012}, but AO-assisted imaging is currently the only option to reach both the diffraction limit of the telescope and provide a  contrast below $10^{-6}$ at a few resolution elements. The size of a resolution element, e.g. the angular resolution, is given by the size of the telescope pupil. For of a circular aperture, the point-spread function (hereafter PSF) is an Airy function of full width at half maximum $\sim \lambda/D$ where D is the diameter of the telescope and $\lambda$ the wavelength. The first three lines of Table \ref{tab_fwhm} summarises the angular resolution of an 8m diffraction-limited telescope in the main optical and near-infrared filters. In the optical, this corresponds to the angular diameter of the more massive nearby stars.

\begin{table}
\caption{Trade-off between angular resolution, AO performance and sensitivity for planets in the different optical and near-infrared filters.}
\label{tab_fwhm} 
\begin{tabular}{p{1.cm} p{4cm} | p{0.8cm} p{0.8cm} p{0.8cm} p{0.8cm} p{0.8cm} p{0.8cm} p{0.8cm}}
\hline\noalign{\smallskip}
\multicolumn{2}{c|}{Band} &	V &	R &	I & 	J &	H & 	Ks &	Lp \\
\noalign{\smallskip}\svhline\noalign{\smallskip}
\multicolumn{2}{c|}{$\lambda$ (\micron)} &	0.55 &	0.65 &	0.82 &	1.22 &	1.63 &	2.2 & 	3.8 \\
\hline
\multicolumn{2}{c|}{Angular resolution (mas)} &	14 &	17 &	21 &	31 &	42 &	57 &	98 \\
\hline
First  & Typical Strehl$^{a}$  (\%) & $<5$ & $<5$ & $<5$ & 5 & 19 & 40 & 73 \\
gen. & Typical contrast$^{b}$ at 0.5\arcsec  ($\times10^{-6}$) & NA & NA & NA & 1.6 & 1.4 & 1.0 & 4.4 \\
AO$^a$ & Corresponding sensitivity$^{c}$  (\Mjup) & NA & NA & NA & 14/43 & 12/37 & 10/32 & 5/12 \\
\hline
\multirow{3}*{xAO$^d$}  & Typical Strehl$^{a}$ (\%) & 26 & 38 & 55 & 76 & 86 & 92 & 97 \\
 & Typical contrast$^{e}$ at 0.5\arcsec ($\times10^{-5}$) & 5 & 4 & 3 & 1.7 & 1 & 0.5 & 0.2 \\
 & Corresponding sensitivity$^{c}$  (\Mjup) & NA & NA & NA & 6/13 & 5/12 & 4/11 & 2/7 \\
\noalign{\smallskip}\hline\noalign{\smallskip}
\end{tabular}
$^a$ 40\% Strehl was assumed at K band, and the Strehl scales as $Sr_{\lambda_2}=Sr_{\lambda_1}^{(\lambda_1/\lambda_2)^2}$.\\
$^b$ a contrast of $1\times 10^{-4}$ was assumed at H band. The scaling in wavelength follows Eq. \ref{eq_contrast_serabyn}.\\
$^c$ The two values refer respectively to a 10 and 100 Myr-old self-luminous planet, orbiting an A0V star. The luminosity to mass conversion used the AMES-COND evolutionary tracks \cite{Baraffe2003}.  \\
$^d$ 92\% Strehl was assumed at K band.\\
$^e$ $1\times 10^{-5}$ was assumed at H band.\\
\end{table}

\begin{figure}[b]
\sidecaption
\includegraphics[width=0.4\textwidth]{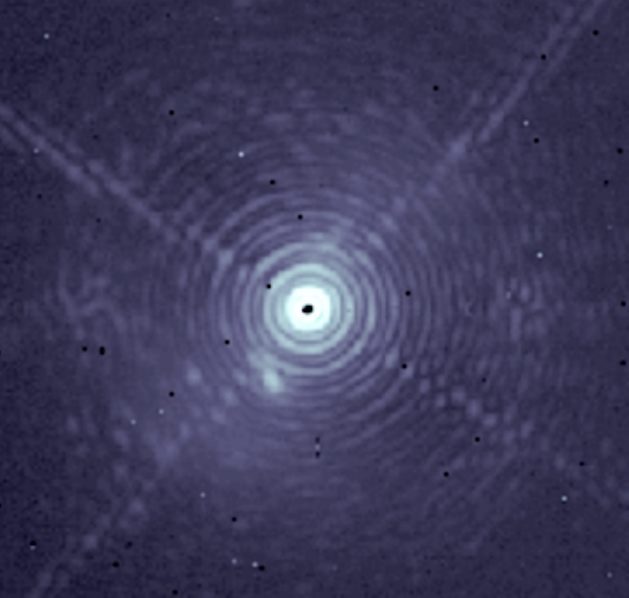}
\caption{Typical high-strehl ($\sim$90\%) raw PSF, as seen from the real-time display during the standard operations of the xAO instrument VLT/SPHERE on a very bright star in the H2 band. The Airy rings are clearly visible, along with the diffraction spikes of the spiders of the telescope and many bright speckles, usually pinned to the diffraction pattern.}
\label{fig_diffraction}       
\end{figure}

To reach the contrast requirements, AO alone is however not sufficient because residual starlight still contaminates the region of interest within 1\arcsec, as illustrated in Figure \ref{fig_diffraction}. These residuals come from both the diffracted light of the telescope entrance pupil, and the residual wavefront error due to uncorrected atmospheric perturbations and imperfect optics in with the telescope and instrument. 
At four resolution elements, the Airy pattern still reaches an intensity of $3 \times 10^{-4}$ the peak value. Therefore, detecting a signal at this level of contrast without any other high-contrast technique is highly ineffective because significant time must be spent to get enough signal. One must rely on additional diffraction light suppression technique, e.g. coronagraphy. 
The second source of residual starlight is more difficult to address. Wavefront errors create speckles in the image plane and mimic point-sources, degrading the contrast \cite{Racine1999}. Unlike the diffraction pattern, speckles vary temporally on different timescales: from a fraction of milliseconds for non-corrected atmospheric speckles to hours or days for quasi-static speckles slowly variable with temperature, mechanical flexions or the rotation of optical parts. Moreover, they can interfere with the diffraction pattern of the pupil and be reinforced to create  pinned speckles \cite{Aime2004}.  Adequate observing strategies and post-processing techniques based on differential imaging can remove the remaining starlight residual to the necessary level. Although coronagraphy and differential imaging are not the focus of this review, they put stringent constraints on the AO system that will be discussed here.

\subsection{From pioneering adaptive optics experiments to extreme adaptive optics system}
\label{sec_history}

AO developments have seen a tremendous progress over the past 30 years, from the first experiments to the most advanced systems now in operations and known as extreme AO systems. The concept of compensating astronomical seeing was first proposed by Babcock in 1953 \cite{Babcock1953}. Military research further developped the concept, and the first prototype for astronomical observations, called COME-ON was installed in 1989 at the ESO/La Silla Observatory \cite{Rousset1990}. ADONIS, an upgraded version of COME-ON and COME-ON+ became the first user-facility instrument equipped with an AO system \cite{Beuzit1994}. Ten years later, VLT/Naos and Keck-AO were the workhorses of the first generation of adaptive optics systems, integrating the lessons learnt from the first pioneering experiments. Combined to the science camera Conica and NIRC2 respectively, these instruments led to many science breakthroughs, such as the first discovery of an exoplanet with VLT/NaCo in 2004 \cite{Chauvin2004}, at a projected separation of 0.8\arcsec, or the discovery of a system of four giant planets orbiting the star HR8799 in 2008 \cite{Marois2008}. In parallel, the design choices of a second generation of AO systems were made, and these extreme AO (xAO) instruments saw their first light in the past few years. The high contrast requirements set new constraints on the AO systems, in order to control the wavefront to an exquisite level, to feed high-rejection coronagraph and to allow differential imaging and maintain the temporal evolution of aberrations as slow as possible. These new systems benefited from the leassons learnt from the first generations of instruments and will in turn provide valuable feedback for the on-going design of AO systems for extremely large telescopes. The first xAO systems on sky were P3000K-P1640 \cite{Oppenheimer2012,Dekany2013}, followed by MagAO \cite{Close2012}. Now three systems had their first light in the last two years: GPI at Gemini South \cite{Macintosh2014}, SCExAO at Subaru \cite{Guyon2010} and SPHERE at the VLT \cite{Beuzit2008}. Table \ref{tab_AO_systems} summarises the main instruments / AO systems contributing or havint contributed to the field. The design of those instruments are a trade-off between angular resolution, AO performance and planet sensitivity, as illustrated by the last rows of Table \ref{tab_fwhm} , that compare the typical AO performance, contrast and sensitivity for both first generation AO systems and xAO systems in different filters.

\begin{table}
\caption{Main instruments benefiting from AO and high-contrast imaging capabilities. This non exhaustive list groups the instruments by generation.}
\label{tab_AO_systems} 
\begin{tabular}{p{4cm}p{2.4cm}p{2cm}p{4.9cm}}
\hline\noalign{\smallskip}
Instrument & Telescope & Wavelength (\micron) & Operations$^a$  \\
\noalign{\smallskip}\svhline\noalign{\smallskip}
ADONIS & La Silla 3.6 & 1-5 & 1996-? \\
PUEO & CFHT & 0.7-2.5 & 1996-2013 \\
\noalign{\smallskip}\hline\noalign{\smallskip}
NaCo & VLT & 1-5 & 2002 \\
Lyot Project & AEOS & 0.8-2.5 & 2003-2007\\
ALTAIR-NIRI & Gemini N. & 1.1-2.5 & 2003 \\
NIRC2 & Keck & 1-5 & 2004 \\ 
NICI & Gemini S. & 1.1-2.5 &  2007 \\
HiCIAO & Subaru & 1.1-2.5 & 2009 \\
\noalign{\smallskip}\hline\noalign{\smallskip}
PALM-3000/P1640 & Palomar 200" & 1.1-1.65 & 2009 \\
FLAO/LMIRCam & LBT & 3-5 & 2012 \\
GPI & Gemini S. & 1.0-2.3 & 2013 \\
MagAO/VisAO & Clay & 0.5-5 & 2014 \\
SPHERE & VLT & 0.5-2.3 & 2014 \\
SCExAO & Subaru & 0.5-2.2 & 2015 \\
\noalign{\smallskip}\hline\noalign{\smallskip}
\end{tabular}
$^a$ Instruments without end date are still in operation.
\end{table} 

\section{Fundamentals of high-contrast adpative optics systems}
\label{sec_key_elements}

\subsection{Characteristics of images distorted by the atmospheric turbulence}

The limit of angular resolution set by the atmopsheric turbulence in the absence of AO correction is $\lambda/r_0$ where $r_0$ is the Fried parameter, scaling as $\lambda^{6/5}$. To evaluate the level of performance of an AO system, the correlation time, also known as the Greenwood time delay is the most relevant parameter. It is defined as $\tau_0=0.314r_0/v$ where $v$ is the mean wind speed weighted by the turbulence profile along the line of sight \cite{Roddier1981}. This parameter sets the speed at which an AO system has to react to correct for the atmospheric turbulence. It is also proportional to $\lambda^{6/5}$, therefore correcting in the near-infrared is easier than in the optical where the turbulence evolves faster.

\subsubsection{Architecure of an AO system}

We review here briefly the architecture and key parameters of an AO system before presenting the specific constraints set by high-contrast observations. We refer the reader to the review \cite{Roddier1999} for further details on general AO systems.

\begin{figure}[b]
\sidecaption
\includegraphics[width=0.5\textwidth]{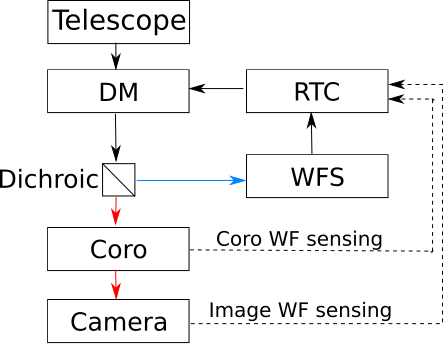}
\caption{Architecture of a standard AO system (plain lines and rectangles). The dashed lines represent the additional features implemented in xAO systems (described in section \ref{sec_NCPA}}
\label{fig_architecture_AO}       
\end{figure}

The architecture ot the first generations of AO systems is composed of three key elements represented in Fig. \ref{fig_architecture_AO}: 
\begin{itemize}
\item a wavefront sensor (WFS), whose role is to measure the optical disturbance in quasi real time; the measurements are sent to
\item a real-time controller (RTC) that reconstructs the wavefront and sends the command to 
\item a deformable mirror (DM) that corrects for the distorted wavefront. 
\end{itemize}

\subsection{Wavefront sensing}

The wavefront sensor is the element measuring in real time the distorsion of the wavefront. Spatial resolution, speed and sensitivity are the three parameters driving the design. An overall review of WFS is given in \cite{Rousset1999}. In the optical and near-infrared, there exists no sensor to measure directly the phase, therefore it must be encoded in intensity variations, and different techniques have been developped. The main challenge consists in getting enough spatial resolution to drive as many modes as the deformable mirror can correct, as fast as the turbulence evolves and without amplifying and propagating the measurement noise. The fundamental limit is set by photon noise.

\begin{figure}[b]
\includegraphics[width=0.9\textwidth]{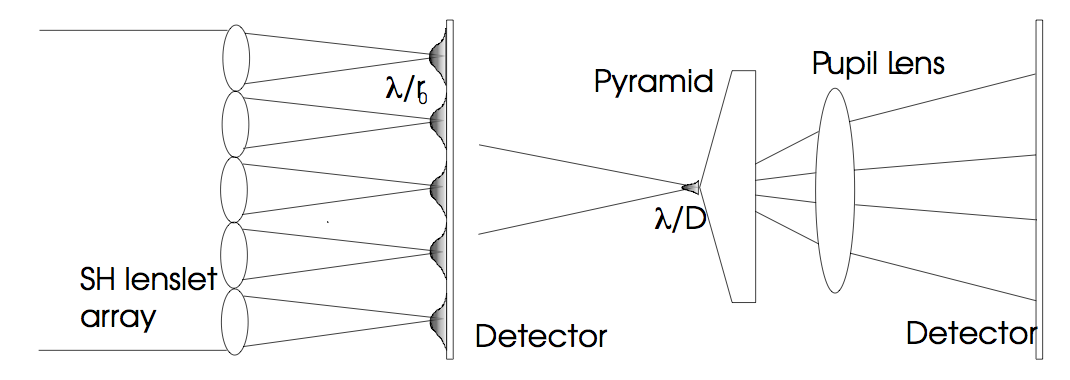}
\caption{Comparison between the Shack-Hartmann (left) and pyramidic (right) WFS, illustrating the highest sensitivity of low order aberrations of the PWFS when a tilt by $\lambda/D$ is introduced in the wavefront (from \cite{Ragazzoni1999}).}
\label{fig_SH_pyr}       
\end{figure}

The two most common implementations measure the slope of the wavefront. For the Shack-Hartmann WFS (SHWFS),  a lenslet array placed in a conjugated pupil plane samples the incoming wavefront (Fig. \ref{fig_SH_pyr} left). Each lenslet creates an image of the source, called a spot, at its focus. In presence of a non-planar incident wave, the lenslet receives a tilted wavefront and the spot is shifted. Therefore, measuring the spot displacement enables to derive the local slope of the wavefront in each lenslet. This requires many pixels to locate accurately the center position of each spot; these are subsequently operating in a low flux regime which requires high sensitivity- and low noise detectors\footnote{Hence the use of Electron Multiplying Charge Coupled Device (EMCCD), for instance in VLT/SPHERE.}. Because this is a relative measurement, it requires also an accurate and regular calibration with a flat wavefront, to avoid any drift. It is a well-proven and mature technology, implemented both by Gemini/GPI and VLT/SPHERE. The pyramid wavefront sensor (PWFS) is a more recent development proposed in 1995 \cite{Ragazzoni1996}. A pyramidal prism is inserted in a focal plane. Each face of the prism deflects the light in a different direction and a lens relay conjugates the four apparent exit pupils onto four pupil images on the detector (Fig. \ref{fig_SH_pyr} right). The prism can be fixed or oscillating. The measured flux in each quadrant can be related to the slope of the wavefront. The PWFS is more sensitive to low-order modes than  the SHWFS \cite{Ragazzoni1999}. This drawback of the SHWFS is of critical importance because many high-rejection coronagraphs require an excellent correction of low-order aberration to provide a high contrast. The PWFS is also less prone to aliasing effects than the SHWFS, although this can be mitigated by the use of an adjustable spatial filter, as in Gemini/GPI and VLT/SPHERE.
The curvature WFS (CWFS) uses two out of focus measurements to derive the curvature (second derivative) of the wavefront \cite{Roddier1988}. It can be implemented by a oscillating membrane. The error propagation is worse than for the SHWFS for equivalent photon noise properties.

The Zernike phase contrast WFS implements an idea developed by Zernike to convert the phase variation in the entrance pupil in an intensity variation in a re-imaged pupil by inducing a phase shift of $\pi/2$ over a diffraction-limited spot in the focal plane \cite{Bloemhof2003}. It is very attractive because it directly converts the phase into intensity instead of measuring the first or secondary derivative of the phase as in the techniques described above, reducing therefore the computation cost and the error propagation. Moreover, the SHWFS is insensitive to certain phase aberration pattern (waffle mode, differential piston), whereas the ZWFS is free from these artefacts.

\subsection{Deformable mirror technologies}

The correction of the wavefront is generally achieved in two or more stages. The tip/tilt mirror corrects for overall shifts of the PSF due either to atmospheric variations, wind or vibrations in the instrument. The correction of higher-order modes of the turbulence is done with a deformable mirror (DM), having up to several thousands actuators. The design of a DM is a trade-off between fast response, density of the actuators, and amplitude and accuracy of their stroke. They are several technologies available, reviewed in \cite{Madec2012}. Some systems even combine different technologies in multi-stage DMs.

Stacked array DMs are made of discrete actuators in ferroelectric material (either piezoelectric or electrostrictive). Lead-zirconate titanate (PZT) and lead magnesium niobate (PMN) are the most common form. When an electric field is applied, it elongates in the direction of the field (longitundinal effect). Although they require high voltages, they provide a large stroke, a high stiffness, reliability and accuracy. This technology was implemented in SPHERE. To increase further the density of actuators or reduce the DM size, MEMS (Micro-Electro Mechanical Systems) are an appealing alternative that use a thin mirror membrane attached to an intermediate flexible support actuated by electrostatic or electromagnetic fields. The use of surface micromachining technologies to etch the electrodes enable to make very compact, high-density and cost-effective DMs. This solution was preferred for Gemini/GPI.
Bimorph DMs exploit the transverse effect of a piezoelectric material. Only two wafers of piezoelectric materials separated by an electrode are needed, which makes manufacturing easier than stacked arrays. They also need high voltages but provide large stroke at a reasonable price. At the VLT, the instruments SINFONI, CRIRES and the Unit Telescopes in interferometric mode all implement this technology.
To provide the large strokes required to drive adaptive secondary mirrors (ASM), voice coil is the preferred solution. The actuators are made of a dense array of voice calls that create a magnetic fields to drive the magnets attached to the mirror thin shell. It provides a fast response but dissipates much heat. It currently equips the LBT and MagAO. 

\section{The transition to extreme AO systems}

Extreme AO systems is an evolution of the concept of single-conjugated AO systems to reach the requirements described in section \ref{sec_requirements} for the detection of exoplanets: high contrast at short separation below 1\arcsec. 
This translates into a requirement for very low wavefront errors, described in section \ref{sec_wfe_xao}. Two approaches are explored in parallel. On the one hand, a lot of effort is devoted to measure and control the wavefront to an exquisite level. To do so, xAO systems use advanced sensing schemes combined with fast, high-density DMs to correct for more than a thousand modes (section \ref{sec_corono} and \ref{sec_NCPA}), filtered by the real-time calculator to match the WFS sensitivity and DM response. Table \ref{tab_XAO_systems} summarizes the main DM and WFS technologies implemented in current xAO systems. On the other hand, a calibration of the static and slowly variable aberrations (section \ref{sec_strategy}) is implemented through specific observing strategies and data reduction algorithms.

\begin{table}
\caption{Extreme AO characteristics \cite{Kasper2012,Dekany2013}}
\label{tab_XAO_systems} 
\begin{tabular}{p{3cm}p{5cm}p{1.5cm}p{1.5cm}}
Instruments & DM technology & WFS & Frame rate (kHz) \\
\hline\noalign{\smallskip}
PALM-3000/P1640 & Electrostrictive, 241 + 3388 act. & SHWFS & 2 \\
FLAO/LMIRCam & Voice coil 672 act. & PWFS & 0.9 \\
MagAO &  Voice coil 585 act. & PWFS  & 1\\
GPI &  Piezoelectric 100  act. + MEMS 1500 act. & SHWFS  & 1.2 \\
SPHERE &  Piezoelectric 1377 act. & SHWFS & 1.2 \\
SCExAO &  Bimorph 188 act. + MEMS 1000 act. & PWFS \\
\end{tabular}
\end{table}

\subsection{Wavefront error requirement for high contrast}
\label{sec_wfe_xao}

The performance of an AO system is quantified using the Strehl ratio, $Sr$, which is the ratio of the peak on-axis intensity of an aberrated wave, to that of a reference unaberrated wave. It can be approximated by the Marechal expression \cite{Marechal1947,Mahajan1983}:
\begin{equation}
Sr = e^{-\sigma_{\phi}^2}
\label{eq_strehl}
\end{equation}
where $\sigma_{\phi}^2$ is the variance of the phase aberration across the pupil and can be decomposed as $\sigma_{\phi}=\frac{2\pi\delta}{\lambda}$ with $\delta$ the root-mean-square wavefront error in nm. As an example, for a wavefront error of 1rad$^2$ rms, which corresponds to $\delta=100$nm at $\lambda=630$nm, the Strehl is 27\%, meaning that 27\% of the PSF energy is in the core of the PSF. The fraction of the PSF energy which is not controlled is critical in high-contrast imaging. Extreme AO systems are systems able to achieve Strehl ratios above 90\%, while standard first generation AO systems typically reach 40\% to 60\% at 1.6\micron, corresponding to $\sim 50$ nm rms in the first case and 200nm rms in the second case. 
Not only is the total variance of the phase $\sigma_{\phi}^2$ critical but the structure of the wavefront error is also important. Wavefront errors at low spatial frequencies need to be controlled at the best level because they correspond to short-separation aberrations in the focal plane. 

As an illustration, let us consider a single mode of the aberrated wavefront in the form of a pure sinusoidal wave wavefront error of amplitude $h$ (expressed in m) and spatial frequency $f$. It can be shown that if the amplitude of the aberration is small with respect to the wavelength, e.g. $h\ll \lambda$, such a mode creates two symmetric replicas of the central PSF at an angular distance $f\lambda$ from the central PSF and with a brightness ratio of $\left( \frac{\pi h}{\lambda}\right)^2$ \cite{Guyon2005}. As an application, with $h=\frac{\lambda}{3140}$, the contrast between the PSF replicas and the central PSF is already $10^{-6}$.

\begin{figure}[b]
\sidecaption
\includegraphics[width=0.62\textwidth]{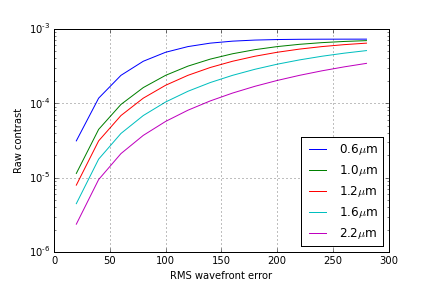}
\caption{Conversion between wavefront error and contrast at different wavelengths in the optical and near-infrared. The typical wavefront error of an xAO system is less than 100nm}
\label{fig_contrast_vs_wfe}       
\end{figure}

In practice the spatial structure of the wavefront error, e.g. its power density function, is unknown and $\sigma_{\phi}^2$ corresponds to an average over many incoherent modes of the aberrated wavefront. Relating the level of aberration $\sigma_{\phi}^2$ to the science requirements of exoplanet detection, e.g. the contrast, is not straight-forward. \cite{Serabyn2007} provide a rough estimate of the level of contrast reached in the well-corrected area of a high-order deformable mirror:


\begin{equation}
C = \frac{1-Sr}{N_{act}}
\label{eq_contrast_serabyn}
\end{equation}
where $N_{act}$ is the number of actuators of the deformable mirror. Combining Eq. \ref{eq_contrast_serabyn} with Eq. \ref{eq_strehl} yields the following expression for the contrast as a function of the rms wavefront error:
\begin{equation}
C = \frac{1-e^{ -  \left( \frac{2 \pi \sigma}{\delta} \right)^2}}{N_{act}}
\label{eq_contrast}
\end{equation}
It is represented in Fig. \ref{fig_contrast_vs_wfe}  for typical wavelengths in the optical and near infrared. Because the rms of the phase error $\sigma_{\phi}^2$ scales with $\lambda/D$, the control of the wavefront error is less critical for AO systems working in the near-infrared than in the optical. 

\begin{figure}[b]
\includegraphics[width=0.9\textwidth]{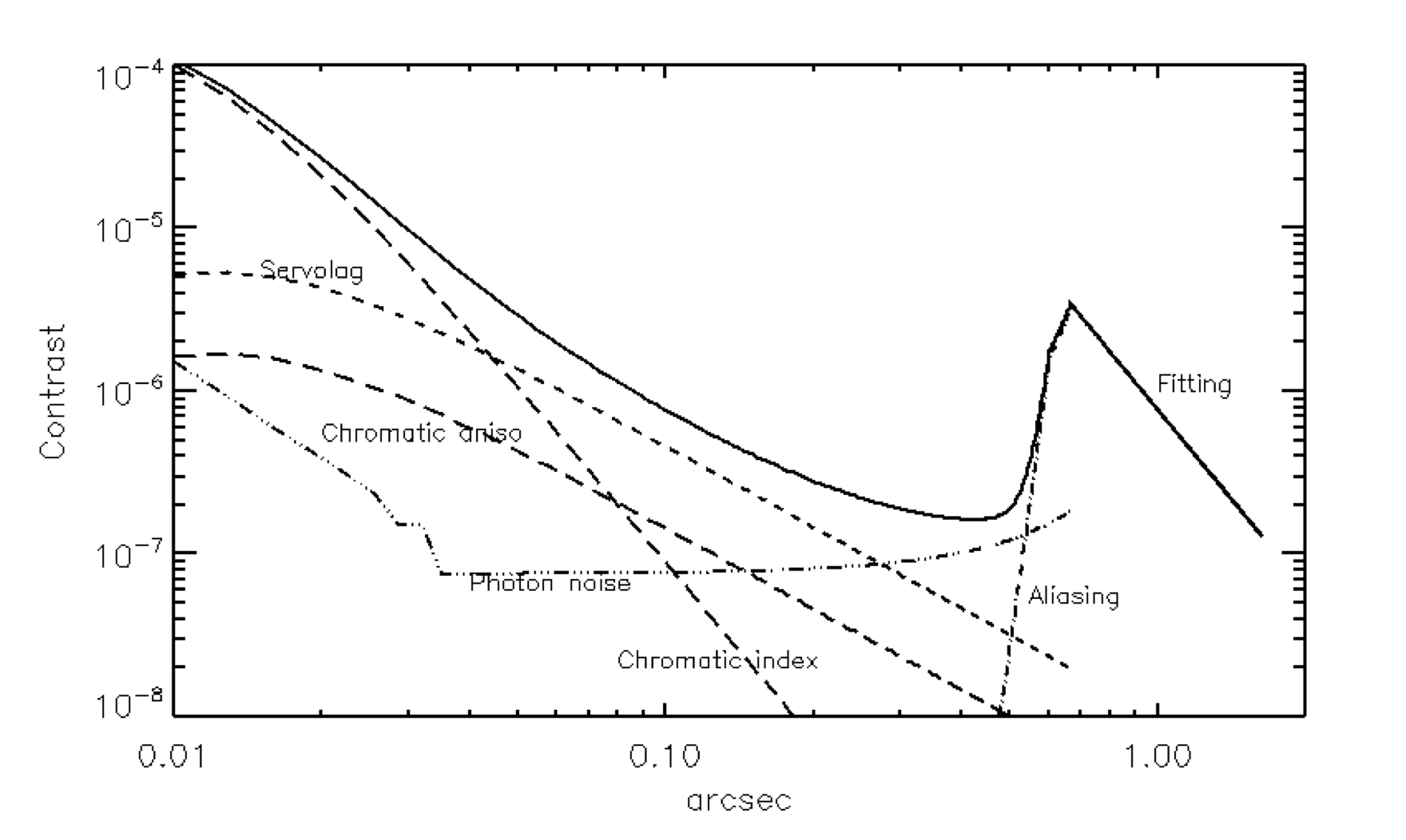}
\caption{Typical error budget as a function of angular separation for an extreme AO system running at 3kHz on a bright (I=6) star, at 1.6\micron{}, with a fixed pyramid WFS, from \cite{Kasper2012}}
\label{fig_ao_error_contribution}       
\end{figure}

In AO-assisted imagers with fast WFS, the major contributor to the error in the wavefront correction is the fitting error.  It is due to the finite number of actuators of the DM, and scales with $N^{-5/6}_{act}\left( \frac{D}{r_0} \right)^{5/3}$. The cutoff frequency which corresponds to the largest spatial frequency that can be corrected by the DM, is given by $f_c = \frac{N\lambda}{2D}$ where N is the number of actuators on a side. In high contrast imaging, the goal is to reach the deepest in-band contrast, within the well-corrected area of the DM, e.g. for $f<f_c$. As detailed in \cite{Kasper2012} and \cite{Guyon2005}, the breakdown of the error within this well-corrected region is strongly dependent on the design choices of the AO system, especially the WFS concerning the associated photon noise. We provide in Fig. \ref{fig_ao_error_contribution} the typical error budget expressed in contrast as a function of separation for a 30m-telescope.
Servolag is the significant contributor between 0.05\arcsec and 0.3\arcsec. It is due to the finite temporal bandwith of the AO, limited by the frequency of the WFS. It scales with $\left( \frac{\tau}{\tau_0} \right)^{5/3}$ where $\tau$ is the time lag in the AO loop and $\tau_0$ the atmospheric correlation time. Last, at very short separation below a few resolution elements, the chromaticity of the optical path length difference and the amplitude aberrations induced by scintillation start to dominate the error budget. 

\subsection{Coronagraphy and diffraction control}
\label{sec_corono}

The role of a coronagraph is to block the starlight and let as much off-axis signal (planet or circumstellar material emission) as possible through the system. It can only remove the coherent, static part of the diffraction pattern but cannot remove speckles due to wavefront errors. This is still greatly valuable because it reduces the photon noise of the diffraction pattern and the coherent amplification between the speckles and the diffraction pattern described in \cite{Aime2004} and visible in \ref{fig_diffraction}. In practice, detectors have a limited dynamical range, therefore the use of coronagraphs avoids saturation and detrimental bleeding. It also limits scattering and parasitic reflexions in the optical train, downstream of the coronagraph. 

Most coronagraph designs are a trade-off between coronagraphic rejection, throughput, inner working angle (IWA) and angular resolution. The state-of-the-art designs are reviewed in \cite{Guyon2006} and more recently in \cite{Mawet2012}. They can be sorted between amplitude masks and phase masks, whether they act on the ampitude or the phase of the wavefront, or between focal and pupil masks, wether they are located in a pupil or in a focal plane.  

\begin{figure}[b]
\includegraphics[width=\textwidth]{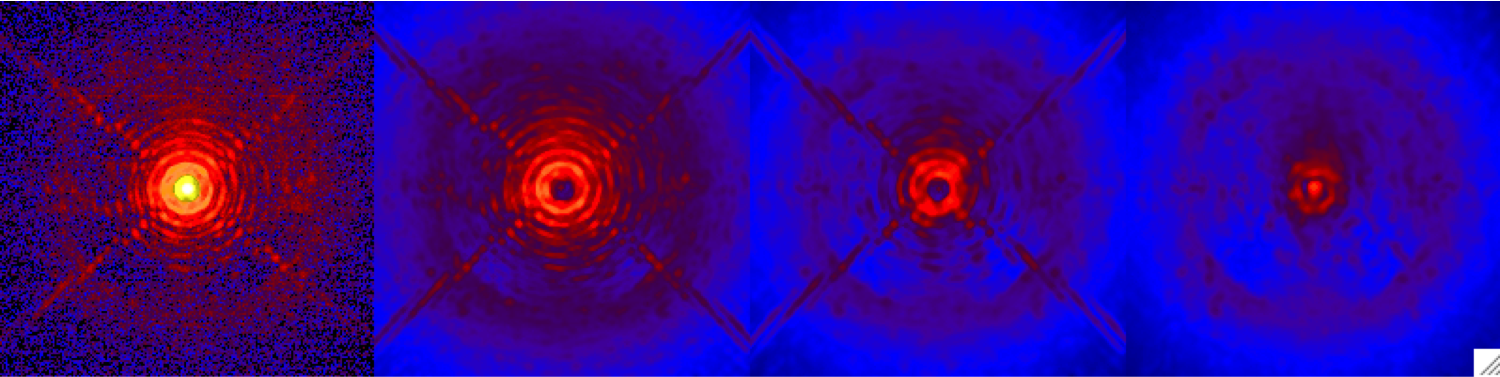}
\caption{Typical high-strehl raw PSF in the presence of turbulence in the VLT pupil (circular aperture obscured by the secondary mirror held by the spiders): without coronagraph (left), with a Lyot stop (middle left), with a Lyot stop  and apodizer (middle right), with apodizer, Lyot stop and pupil stop masking the telescope spiders (right). These images were obtained with SPHERE \cite{Beuzit2008}, the well-controled radius measures 0.8\arcsec.}
\label{fig_diffraction_control}       
\end{figure}

In VLT/SPHERE, Gemini/GPI and Subaru/HiCIAO, apodized Lyot coronagraphs are mostly used. They represent an evolution of the Lyot coronagraph to include an apodized entrance pupil to further improve the achievable contrast \cite{Soummer2003} by removing the diffraction pattern. Fig. \ref{fig_diffraction_control} illustrates the level of light suppression and diffraction control that can be done by combining an apodizer, a Lyot stop and a pupil stop. They are typically limited to an IWA of $3-4\lambda/D$ in their current design. 

To provide a smaller IWA than conventional Lyot coronagraphs, \cite{Guyon1999} proposed the concept of a focal plane phase mask. The four-quadrant phase mask \cite{Rouan2000} and vortex coronagraph known as the AGPM \cite{Mawet2005} are evolutions of this concept and are used today on VLT/NaCo. The latter has been successfully commissioned on Keck/NIRC2, on VLT/VISIR in the N band and is foreseen to equip VLT/SPHERE. They enable a smaller IWA, but chromaticity is hard to achieve.
Pupil apodization is another technique developped in order to smooth the pupil edges that create the Airy rings, implemented through a continuous amplitude mask or a binary mask known as a shaped pupil \cite{Kasdin2003}. Loss-less pupil apodization can be achieved through phase remapping. Successful implementations of this concept include the apodizing phase plate (APP) on VLT/NaCo \cite{Kenworthy2007} or the phase induced amplitude apodization coronagraph (PIAA) on Subaru/SCExAO \cite{Guyon2003}.

\subsection{Low-order wavefront sensing and non-common path aberrations}
\label{sec_NCPA}

Most coronagraphs require the best possible correction of low-order aberrations, most importantly tip/tilt, focus, astigmatism and coma, in order to suppress light efficiently at their IWA. Uncorrected low-order wavefront error create light leaks around the coronagraph mask that mimic point-sources and degrade the contrast. 
This requirement is challenging in standard AO systems because there are non-common path aberrations (NCPA) between the WFS arm and the science arm and the WFS might not be sensitive enough to these low-order aberrations, as it is the case for the SHWFS. 
A number of practical solutions exist for current xAO systems, they are reviewed in \cite{Mawet2012}. In short, the simple architecture of Fig. \ref{fig_architecture_AO} (plain lines) is modified to include two additional feedbacks from the coronagraphic plane and image plane in order to sense the wavefront errors close to the coronagraph and at the same wavelength as the science camera. The first feedback loop requires a dedicated calibration system. In VLT/SPHERE, this is implemented through a beamsplitter close to the coronagraphic focus, sending a small fraction of the light to the differential tip/tilt sensor \cite{Petit2014}. A distinct choice was made by Gemini/GPI and Subaru/SCExAO where a modified Mach-Zehnder interferometer combines the light reflected off the coronagraph spot with a reference wavefront. 
The second feedback uses directly the image recorded from the science camera to reconstruct the NCPA using a phase diversity algorithm \cite{Sauvage2011,Paul2013}. 

\subsection{Observation strategies for improved stability and speckle removal}
\label{sec_strategy}

\subsubsection{Stability considerations}

\begin{figure}[b]
\sidecaption
\includegraphics[width=0.62\textwidth]{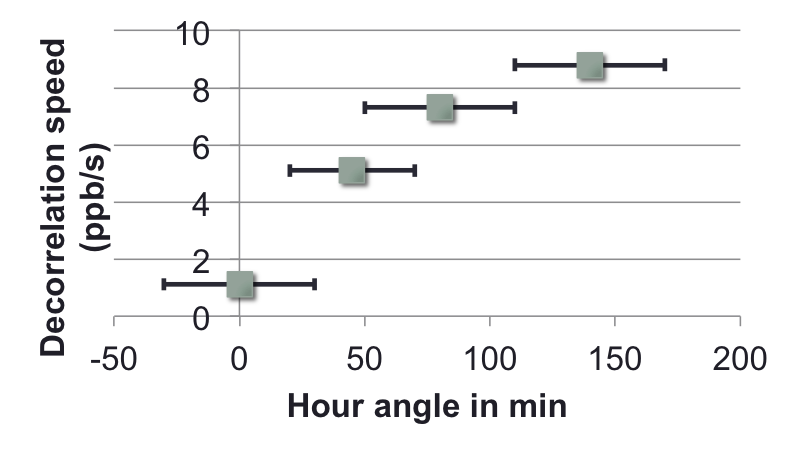}
\caption{Decorrelation speed as a function of the hour angle. These data were obtained with NaCo in the Lp band and averaged over one hour (horizontal error bar) to derive meaningful conclusions (Milli et al. in prep). Decorrelations is much slower around meridian than at larger hour angles.}
\label{fig_decorrelation_speed}       
\end{figure}

Soon after the first generation of AO systems came online, it was realized that quasi-static residual speckles were the largest source of noise at short separation that prevent detection from faint companions \cite{Racine1999}. Unlike atmospheric residuals that would average over time, these speckles are long-lived \cite{Hinkley2007}. They come from mechanical flexures, imperfect optics, non-common path errors between the WFS arm and the science arm that are slowly varying with the tracking of the telescope and rotation of the optics. Because most of these slowly varying wavefront errors are fixed in a pupil frame rather than in a sky frame, high-contrast imaging is performed in pupil-stabilized mode. This way, the PSF is kept as stable as possible during the science observations and can be calibrated in the post-processing stage. On the Cassegrain focus of an alt/az telescope (e.g. Gemini/GPI), this means turning off the rotator, whereas on the Nasmyth platform (e.g. VLT/SPHERE), this means introducing a derotator as early as possible in the optical train. Around meridian passage, the apparent motion of a star on sky is the slowest, the altitude motion of the telescope goes to zero. This stable configuration can be traced down to the science camera, as shown in Fig. \ref{fig_decorrelation_speed}, showing that deep coronagraphic images decorrelate slower around meridian passage. 

This is interesting to note that the two xAO instruments VLT/SPHERE and Gemini/GPI have adopted different solutions to minimise the impact of slowly variable aberrations. GPI is a light and compact instrument attached to the moving Cassegrain focus, it implements a calibration unit (see section \ref{sec_NCPA}) to probe in real time the evolution of the quasi-static speckle and control them. On the other hand, SPHERE is a heavy instrument that rests on the Nasmyth platform via an actively-controlled support to damp vibrations to stay as stable as possible.

\subsubsection{Observing strategies}

Once all possible efforts have been made to keep the wavefront error as low as possible and the PSF as stable as possible, the post-processing stage aims at further enhancing the detection of astrophysic signal and removing instrumental speckles. This step relies on the introduction of diversity between the astrophysic signal and the speckles. 

\begin{figure}[b]
\sidecaption
\includegraphics[width=0.62\textwidth]{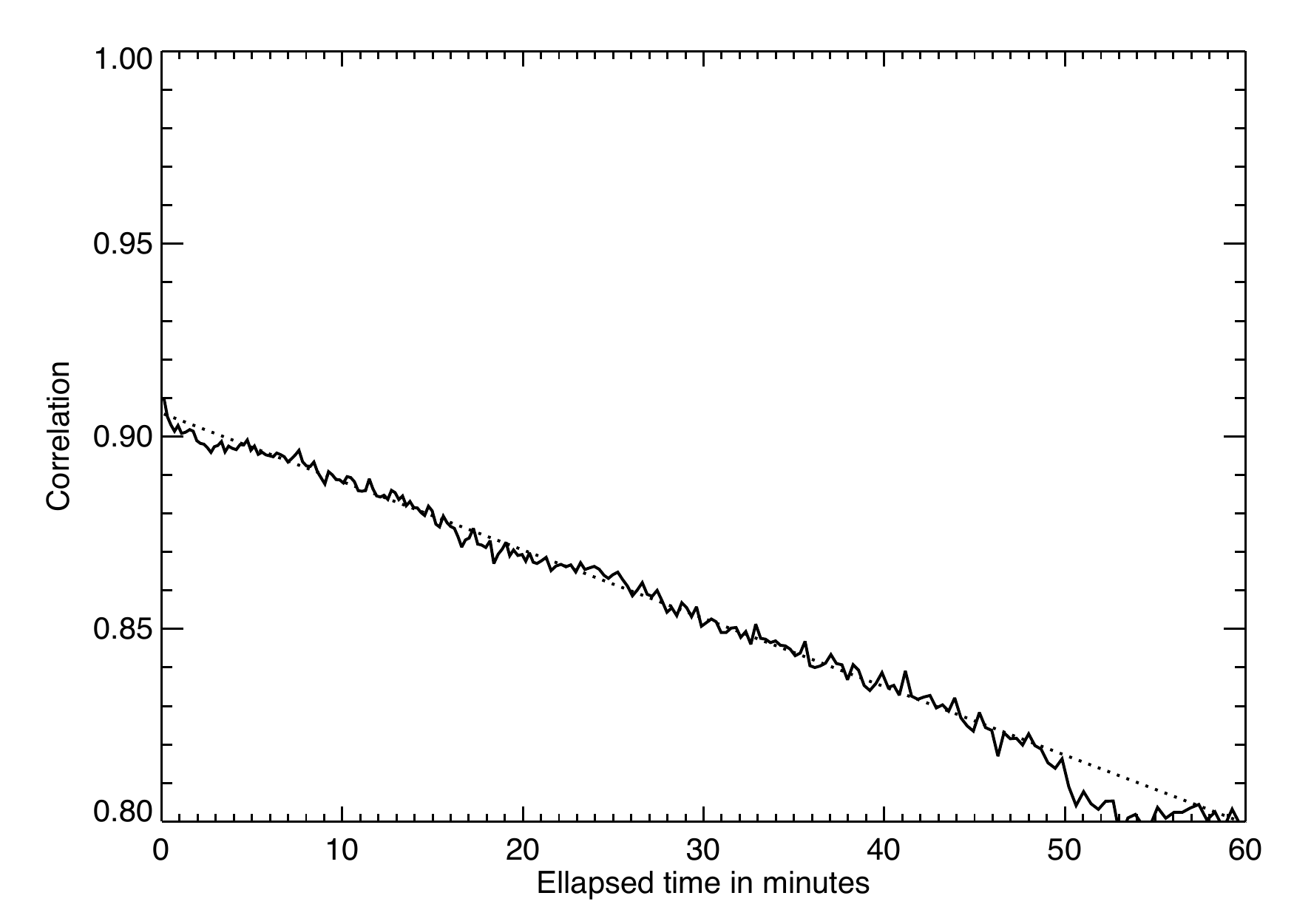}
\caption{Decorrelation over time for a set of deep H band observations with VLT/NaCo. The decorrelation shows a sharp decrease within the first seconds and decreases linearily in the next hour}
\label{fig_decorrelation_vs_time}       
\end{figure}

In pupil-stabilized observation, now the baseline for most high-contrast observations \footnote{except for polarimetry where high polarimetric accuracy may drive the need for a different stabilisation scheme.}, the relative motion of an on-sky signal with respect to the pupil is used to disentangle between a fixed speckle and a rotating companion, a strategy called Angular Differential Imaging (ADI) \cite{Marois2006}. Advanced data reduction algorithms have been developped to further improve the efficiency of ADI \cite{Lafreniere2007,Mugnier2008,Soummer2012,Amara2012,Galicher2011}. 
However, is intrinsically limited by two considerations: \\
1) the decorrelation of the PSF over time \cite{Hinkley2007}, illustrated in Fig. \ref{fig_decorrelation_vs_time}, linear on a timescale of several tens of minutes but very steep within the first seconds \\
2) the very slow rotation of the field at very short separations where planet are expected, which implies a lot of self-subtraction of the astrophysical signal, especially for the case of extended sources such as disks \cite{Milli2012}. \\

Reference star differential imaging (RDI) is a solution to circumvent these intrinsic difficulties. It implies observing a reference star as close as possible in time and space, ideally with the same parallactic angle variation to keep the motion of the optics as close as possible to the science observations. In practice, fast switching between science targets and calibrators suffers from the overheads of the WFS acquisition, but solutions exist to close the loop after switching target without reacquiring on the WFS, as implemented on VLT/NaCo ("star hopping" \cite{Lacour2011,Girard2012}). On the other hand, it is also possible to rely on a large library of PSF acquired as part of a survey on many stars along several nights or weeks to build an optimal PSF for the star subtraction. Such a strategy was proposed on the Hubble Space Telescope \cite{Soummer2014,Choquet2014} and observing programs are currently on-going to validate this strategy on ground-based instruments in the new high-strehl regime opened by xAO instruments. 

\begin{figure}[b]
\includegraphics[width=0.7\textwidth]{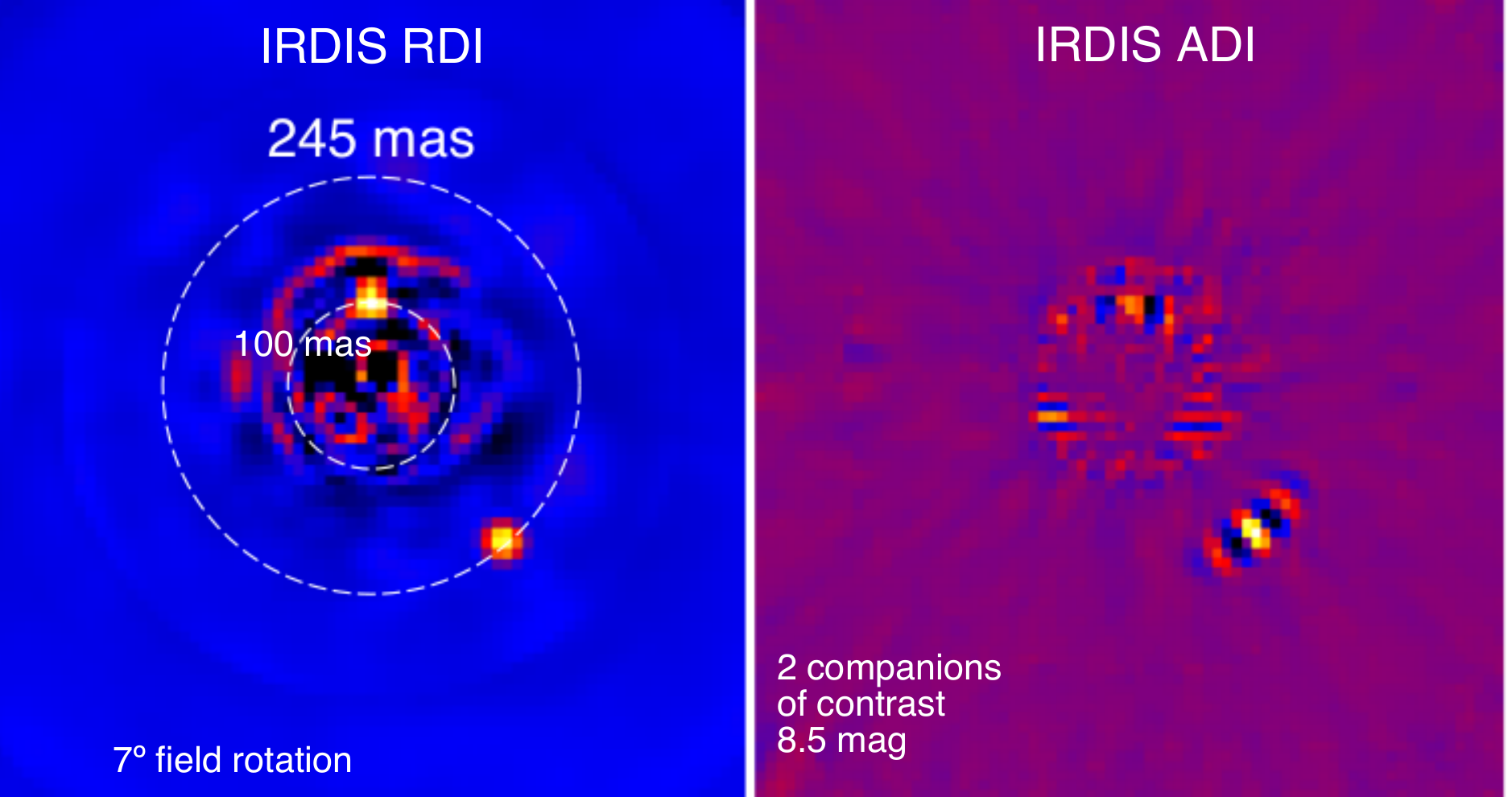}
\caption{Illustration of the power of RDI with VLT/SPHERE in the J-band. A sub-stellar companion of contrast 8.5 magnitude is detected at 0.245\arcsec with a higher SNR in RDI (left image) than ADI (right image). The same conclusion can be drawn for the fake companion injected at 0.1\arcsec. The RDI does not suffer from the self-subtraction issue that create negative side lobes around the companion in ADI. The ADI image was scaled to account for the self subtraction which decreases radially. The field rotation of $7^\circ$ corresponds to only $0.4\lambda/D$ at 0.1\arcsec.}
\label{fig_RDI_power}       
\end{figure}

Simultaneous differential imaging represents another solution, whether spectral (SDI) or polarimetic (PDI). For that, the target needs to have a spectral feature (usually a methane feature for gas giants) or must show linear polarisation (disks or planets in scattered light). SDI relies on the fact that speckles scale with the wavelength unlike on-sky signal, whereas PDI relies on the fact that the thermal emission from the central star is unpolarized unlike scattered light from circumstellar material. Table \ref{tab_diff_imaging} summarises the benefits and drawbacks of each differential imaging technique. Note that ADI can be carried out on top of SDI or PDI.

\begin{table}
\caption{Overview of the most common differential imaging techniques}
\label{tab_diff_imaging} 
\begin{tabular}{p{1cm}p{1.5cm}p{4cm}p{5cm}}
Name & Diversity parameter & Strengths & Drawbacks \\
\hline\noalign{\smallskip}
ADI & pupil/fied relative rotation & easy implementation & PSF decorrelation over time, self-subtraction \\
RDI & science/ref star & no self-subraction & rapid switching required or large library, subject to change in the PSF shape/ AO correction, overheads due to the calibrator \\
SDI & wavelength & simultaneous difference & relies on spectral features, chromaticity \\
PDI & linear polarisation & simultaneous difference, achromatic & relies on linear polarisation feature, calibration of instrumental polarisation \\
\end{tabular}
\end{table}

\section{Science highlights and new challenges}
\label{sec_science}

We provide in this section a few selected examples that showcase the new possibilities offered by xAO instruments. 

\subsection{Disks at very short separations}

\begin{figure}[b]
\includegraphics[width=1\textwidth]{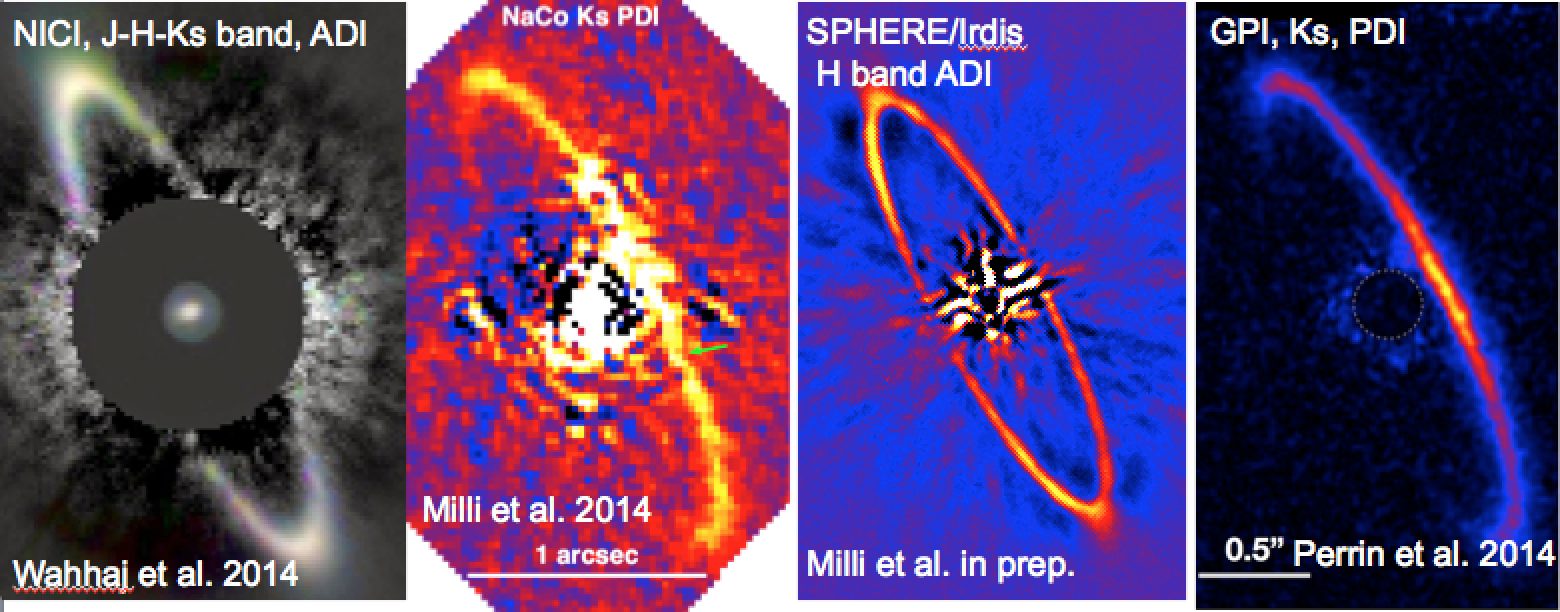}
\caption{Comparison between the first generation AO imagers (left images) and xAO imagers (right images). The first and third images are coronagraphic images obtained in ADI, while the second and fourth are polarized images.}
\label{fig_comparison_HR4796}       
\end{figure}

HR4796 is a prototypical debris disc that makes an ideal benchmark to compare instruments because it has been observed with almost all high-contrast instruments. 
The disc has a semi-major axis of $\sim1$\arcsec and a semi-minor axis of $\sim0.2$\arcsec. It was not detected in early AO imaging with the pioneering AO system COME-ON-PLUS \cite{Mouillet1997}. First generation AO systems revealed the ring along its semi-major axis (see for instance the recent Subaru NICI image from \cite{Wahhaj2014} in Fig. \ref{fig_comparison_HR4796} left). They were mostly  blind to the semi-minor axis until the recent NaCo polarimetric image could reveal the disc almost entirely (Fig. \ref{fig_comparison_HR4796}, second image from \cite{Milli2015}). An uncontrolled mode of the DM (waffle) creates a four-point pattern at the DM cutoff frequency ($7\lambda/D$). 
With the xAO instruments VLT/SPHERE and Gemini/GPI, the disc is detected directly in raw frames of a few seconds without any star subtraction.
The disc was observed in ADI in the H band during VLT/SPHERE commissioning (ESO press release 1417, third image in Fig. \ref{fig_comparison_HR4796}) under poor conditions. Light leak around the coronagraph below $0.2$\arcsec and the small field rotation of only $22^\circ$ prevent a clear detection of the semi-minor axis but the improvement in achieved separation and image quality is striking. 
With GPI, the disc is detected unambiguously in polarimetry in the Ks band (right image, \cite{Perrin2014}). The concept of integral field polarimetry implemented in GPI minimizes differential wavefront error between the two polarization channels, whereas the Wollaston prism of NaCo introduces differential aberrations. 

\subsection{Planets in the optical}

\begin{figure}[b]
\includegraphics[width=1\textwidth]{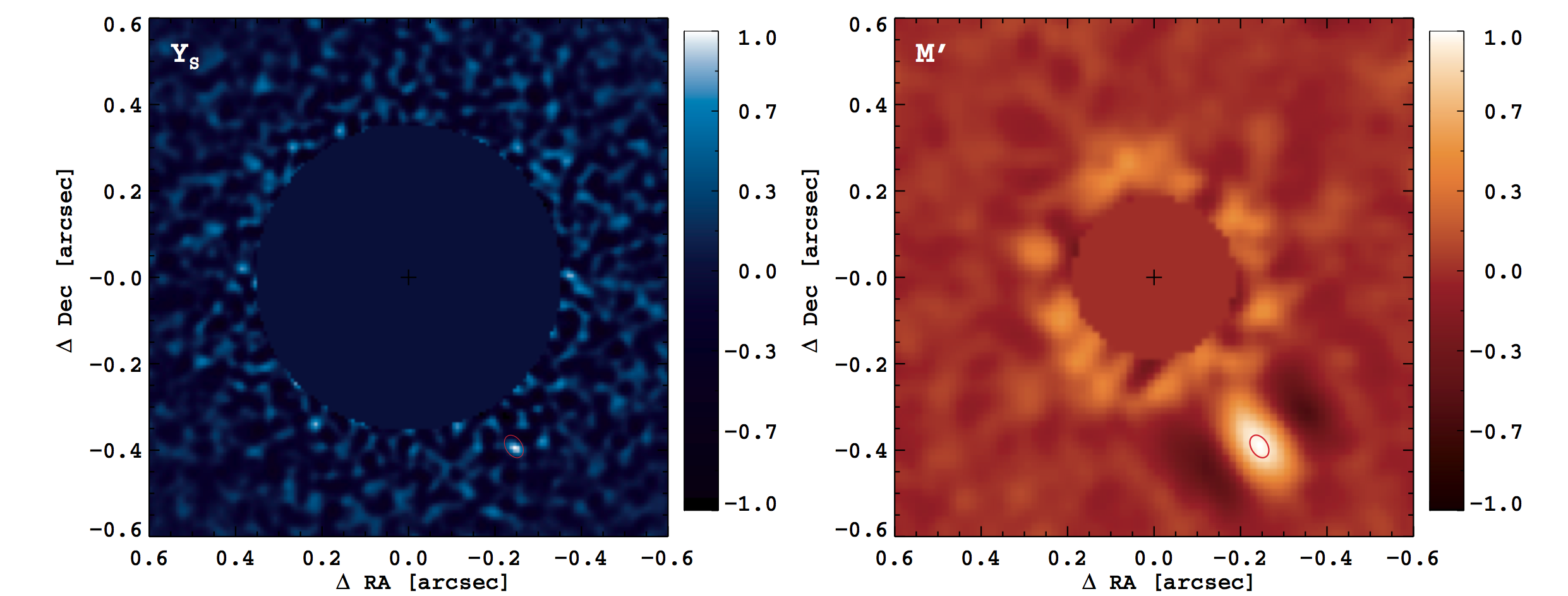}
\caption{Simultaneous detection of $\beta$ Pictoris b in the red optical (left) and in the near-infrared at $M^\prime$ (right) \cite{Males2014}. This illustrates the higher resolution provided by shorter wavelengths.}
\label{fig_bPic_vis}       
\end{figure}

The new generation of xAO systems provides a very high wavefront correction that opens the door to optical AO-assisted imaging, much more demanding due to the smaller wavelengths. The optical range is appealing because the angular resolution is higher, and polarisation can be efficiently used both for starlight rejection and planet/disc characterisation. 
The first system on-sky was MagAO on the Magellan Clay telescope. MagAO is a 585-actuator adaptive secondary mirror with a pyramid WFS \cite{Close2012}. It can operate simultaneously in the optical with the camera VisAO and in the infrared with Clio2. The gas giant $\beta$ Pictoris b was detected for the first time in the red optical (band $Y_S$, that still corresponds to the planet thermal emission) with VisAO in 2012 \cite{Males2014}, with a SNR of $\sim4$ (Fig. \ref{fig_bPic_vis} left) and a Strehl of 40\%.  

Since then, other visible AO instruments arrived on-sky, such as VAMPIRES \cite{Norris2015} fed by the SCExAO system, or SPHERE/Zimpol \cite{Schmid2010}, both implementing differential polarimetry. Zimpol is an unique concept of a high-accuracy polarimeter and imager. The polarimeter is based on the concept of high-frequency modulation of the polarisation to freeze the non-corrected atmospheric residuals, using a ferroelectric liquid cristal operating at 1 kHz. Demodulation is carried out with a dedicated CCD camera synchronized with the modulator and switching the charges alternatively up and down. Combined with the SPHERE xAO system, Zimpol aims at detecting exoplanets in reflected light \cite{Milli2013}, due to its exquisite inner working angle of 30mas ($2\lambda/D$ at 600nm). During the science verification, the Zimpol instrument in imaging mode resolved for the first time the surface of a star: the nearby red giant R Doradus, of diameter 57mas.

\subsection{Spectra of exoplanets and brown dwarfs}

\begin{figure}[b]
\sidecaption
\includegraphics[width=0.5\textwidth]{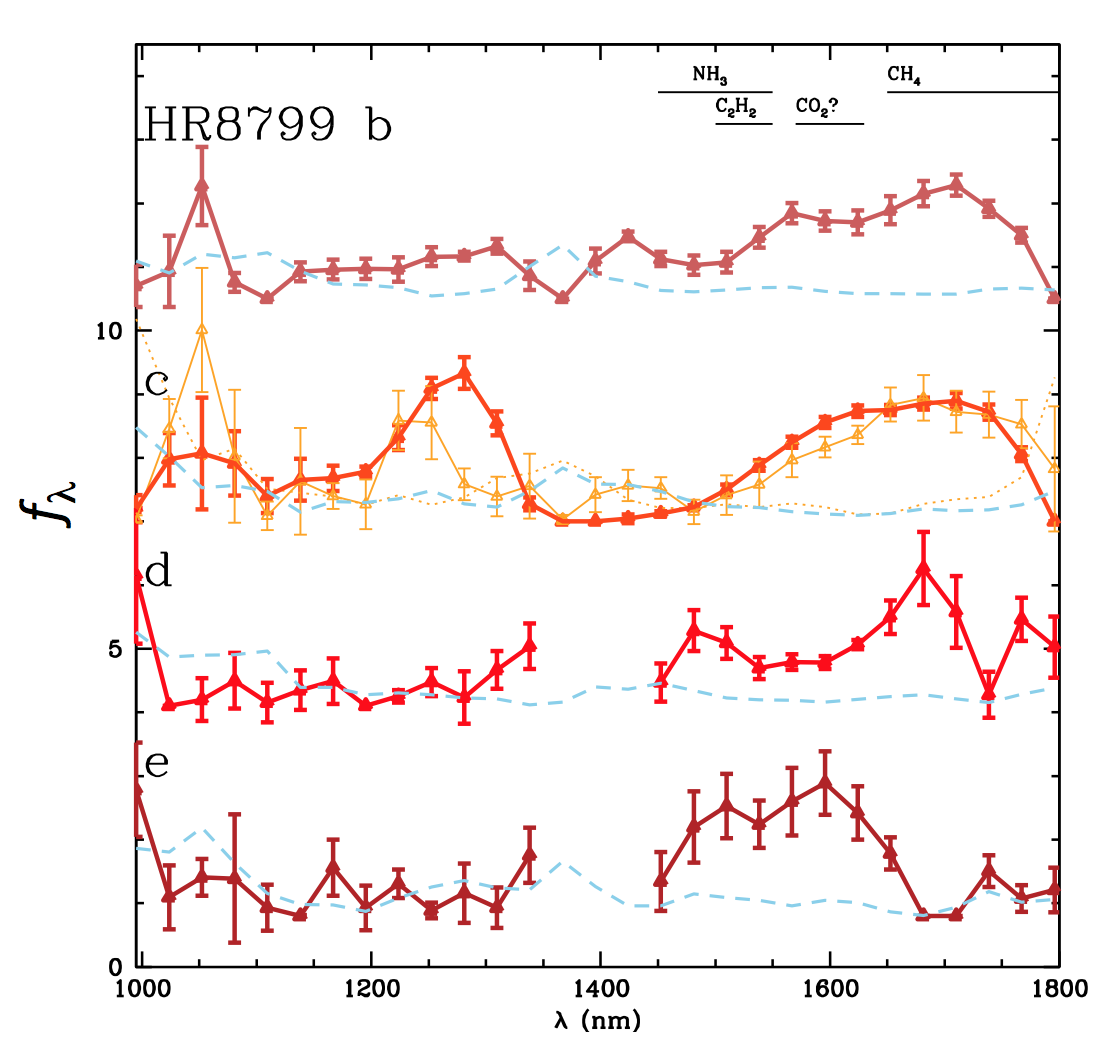}
\caption{Spectra of the four exoplanets detected in the system around HR8799, obtained with a resolution of $\sim35$ with the system PALM3000/P1640 \cite{Oppenheimer2013}.}
\label{fig_spectra_HR8799}       
\end{figure}

Combined to integral field spectrographs (IFS), xAO provided the first spectra of exoplanets. The first IFS on-sky was P1640 at Palomar \cite{Oppenheimer2012}. It revealed the spectra of the 4 known exoplanets HR8799 b, c, d and e (Fig. \ref{fig_spectra_HR8799}) from 995nm to 1769nm, and identified the presence of $CH_4$ along with $NH_3$, $C_2H_2$, and possibly $CO_2$ or HCN in variable amounts in each component of the system. More recently, Gemini/GPI revealed the H band spectra of $\beta$ Picoris b \cite{Chilcote2015} at only 436 mas of the central star. The spectrum, obtained with a resolving power of $R\sim45$, shows a triangular shape, typical of cool low-gravity substellar objects. Extreme AO systems can also be combined to long slit spectroscopy (LSS) to provide medium resolution spectra. The concept of coronagraphic LSS is implemented in VLT/SPHERE \cite{Vigan2008} to enhance starligh rejection. It w`as recently illustrated during Science Verification to reveal  an $R\sim350$ spectrum of the young substellar companion 2MASS 0122-2439B across the YJH bands  \cite{Hinkley2015} .

\section{Conclusions and future challenges}

Today, several large-scale surveys of extrasolar planets using xAO systems are under way. These studies probe a new parameter space at smaller separations and higher contrast and will probably lead to tens of new detections in the coming years, that will help to understand the population and formation mechanisms of giant planets. 
On individual systems, these new capabilities also open up new perspectives. The photometric accuracy will enable to capture temporal variations of exoplanet emission due to non-uniform cloud coverage, or temporal evolution of disk asymmetries as a result of gravitational perturbation or stellar winds. By reaching shorter separations, direct imaging will bridge the gap with the radial velocity technique, to get access to the dynamical mass of giant planets and therefore test the evolutionary models.  

From a technical point of view, the higher Strehl and stability of xAO systems unexpectedly revealed fine instrumental effects that passed unnoticed in first generation systems because of the lack of sensitivity. Such effects include vibration issues, peculiar dome- or low-altitude atmospheric conditions or subtle polarisation mechanisms. It also triggered an intense activity on post-processing techniques and signal detection theory, for which other fields of physics were a major source of inspiration. Observing strategies and data extraction methods valid at large separation and widely used on first generation instruments are not necessarily relevent at very short separation of a few resolution elements and need to be tailored for this new regime. 

These developments and the experience acquired through xAO instruments will contribute to the preparation of the first planet-finder instruments for 30m-class telescopes arriving in the 2020s. Such systems are expected to unveil even shorter separations below 10 mas, but they will also have to overcome new challenges such as chromatic effects or segmented mirrors. Complemented by space-based coronagraphic instruments aiming for deeper contrasts, they are expected to reveal the first images of rocky planets around nearby stars.

\bibliography{}

\end{document}